\documentclass[final,5p,times,twocolumn]{elsarticle}

\usepackage{graphicx}
\usepackage{xcolor}
\usepackage{amssymb}
 \usepackage{amstext}

\usepackage{lineno}

\journal{arXiv}
\begin{document}
\begin{frontmatter}

\title{Raman Signal Reveals the Rhombohedral Crystallographic Structure in Ultra-thin Layers of Bismuth Thermally Evaporated on Amorphous Substrate}


\author[LNCMI,Valencia1]{Carlos Rodr\'\i guez-Fern\'andez}
\author[Leiden]{Kim Akius}
\author[Valencia2]{Mauricio Morais de Lima Jr.}
\author[Valencia1]{Andr\'es Cantarero}
\author[Leiden]{Jan M. van Ruitenbeek}
\author[Leiden]{Carlos Sabater\corref{actualadre}\fnref{label2}} 

\cortext[actualadre]{Present adress: Departamento de F\'\i sica Aplicada and Unidad asociada CSIC, Universidad de Alicante, Campus de San Vicente del Raspeig, E-03690 Alicante,
Spain.}
 \fntext[label2]{Corresponding Author: carlos.sabater@ua.es}
 \address[LNCMI]{Laboratoire National des Champs Magnétiques Intenses, CNRS-UGA-UPS -NSA-EMFL, 25 avenue des Martyrs, 38042 Grenoble, France}
 \address[Valencia1]{Molecular Science Institute, University of Valencia, P.O. Box 22085, 46071 Valencia, Spain}
  \address[Leiden]{Huygens-Kamerlingh Onnes Laboratorium, Leiden University, Niels Bohrweg 2, 2333 CA Leiden, Netherlands}
 \address[Valencia2]{Materials Science Institute, University of Valencia, P.O. Box 22085, 46071, Valencia, Spain}

\begin{abstract}
Under the challenge of growing a single bilayer of Bi oriented in the (111) crystallographic direction over amorphous substrates, we have studied different thicknesses of Bi thermally evaporated onto silicon oxide in order to shed light on the dominant atomic structures and their oxidation. We have employed atomic force microscope, X-ray diffraction, and scanning electron microscope approaches to demonstrate that Bi is crystalline and oriented in the (111) direction for thicknesses over 20 nm. Surprisingly, Raman spectroscopy indicates that the rhombohedral structure is preserved even for ultra thin layers of Bi, down to $\sim 5$ nm. Moreover, the signals also reveal that bismuth films exposed to ambient conditions do not suffer major surface oxidation.
\end{abstract}


\end{frontmatter}


\definecolor{mygreen}{rgb} {0.13, 0.55, 0.13}

\section{Introduction}
The element of the periodic table bismuth (Bi), in bulk, is a fascinating semi-metal, for which the electronic properties and band structure \cite{Hofmann2006,EtranBiSingCryst1963} are unique. Exotic effects such as spin textures and Rashba interactions emerge in this element when its thickness is close to $\sim 20$ nm  \cite{20nmBirashba}. Moreover, when Bi is reduced to a single bilayer oriented in the (111) crystallographic direction (see Fig. 1a)) and is connected by electrodes, the quantum spin Hall effect appears \cite{KaneMele, Murakami}. In other words, for this two-dimensional (2D) material, also called a 2D Topological Insulator (2D-TI) \cite{YoiAndo}, electronic transport is only possible via metallic edge-states. From a technological perspective, 2D-TI are very attractive for applications in spintronics \cite{Wang2016}. 

Measurements of the electronic transport in point contact experiments employing a Scanning Tunneling Microscope (STM) under ambient conditions \cite{BiCsab2013} suggested that a Bi bilayer can be formed between two Bi electrodes after continuous cycles of indentation and retraction. However, a bilayer suspended between electrodes is quite unstable and not suitable for  industrialized processing. For this reason, it is more interesting to explore and characterize Bi deposited on surfaces. While experimental growth of a single layer of Bi in the hexagonal phase has been failed on Si(111) surfaces \cite{FailBi1,Failbi2}, first results were obtained for the growth of a single Bi monolayer on silicon carbide by F. Reis \textit{et al}. \cite{BismutheneSC}. Since electronic devices in industry, and many technical and condensed matter physics laboratory systems, use wafers of silicon with a top layer of silicon oxide (SiO$_{2}$) as electrical insulator it would be very attractive to use these substrates, in order to deposit, grow, or synthesize a bilayer of Bi (111). Currently, we are not aware of any reports that demonstrate that a single bilayer of Bi was grown via thermal evaporated on SiO$_{2}$, however using another expensive and sophisticated approaches as Pulsed Laser Deposition (PLD) the bismuth can be deposited\cite{HaoNPJ2020,Yang2019}. Some studies based in thermal evaporation technique on amorphous substrates show that the crystallographic orientation of the Bi grains are governed by the surface morphology and by the energy of the impinging atoms. For instance, Rodil \textit{et al.} reported that low deposition energies lead to (111)-oriented growth, high deposition energies provide a (012)-orientation, and intermediate energies give random orientations \cite{Rodil2017}.  

\begin{figure}[ht]
\centering
\includegraphics[width=0.99\linewidth]{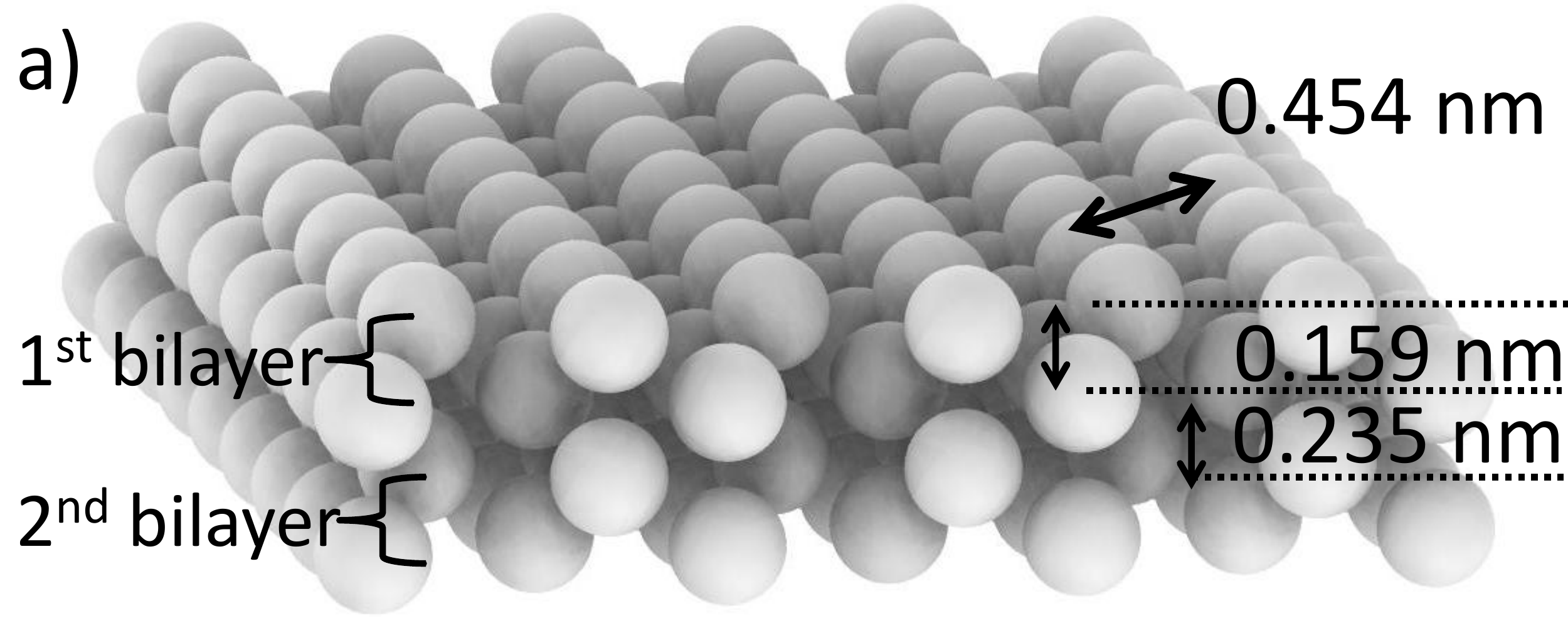}
\includegraphics[width=0.99\linewidth]{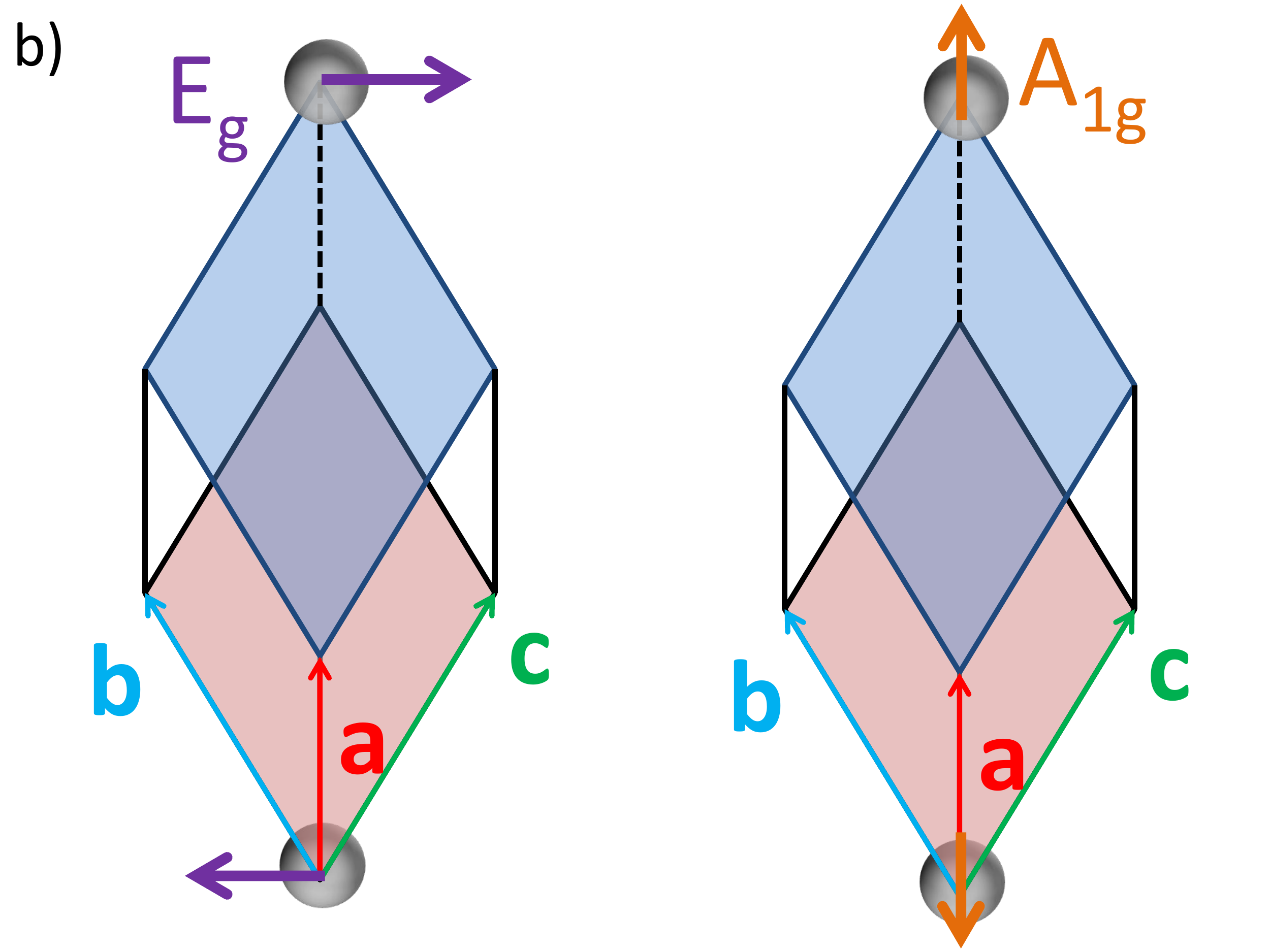}
\caption{a) Illustration two bilayers of Bi(111) and their atomic distances.b) Trigonal unit cell of Bi crystal. Pale blue and pale red rhombs represent the upper and bottom part respectively. Blue, red and green arrows show the unit vectors spanning the lattice. Bi atoms are illustrated by the grey spheres. In the upper and lower atoms, purple and orange bidirectional rows show the modes of vibration labeled as $E_{g}$ and $A_{1g}$.}
\label{trigonal}
\end{figure}

One of the significant challenges for obtaining pure Bi thin films on amorphous substrates is to avoid the oxidation during  the process, otherwise crystallographic orientation and structure are strongly affected. In Bi thin films, this topic is still under debate. There are early reports where Bi thin films become covered by a monolayer of BiO, not by Bi$_{2}$O$_{3}$ Ref. \cite{PRB90BiO}, reports that showed that Bi nanostructures oxidize from the edges and their top surfaces remain unoxidized \cite{Kowalczyk}, or papers that show that Bi (111) has a very limited degradation even after being exposed to air for weeks \cite{Yang2019}. Bismuth oxide exhibits strong polymorphism, consisting of the major part of phases of stoichiometric nature, including $\alpha$-Bi$_2$O$_3$ with monoclinic symmetry, $\beta$-Bi$_2$O$_3$ with tetragonal structure or $\gamma$-Bi$_2$O$_3$ with cubic (bcc) symmetry, among others. By comparison, pure Bi crystallizes in the A7 rhombohedral lattice structure, which belongs to the R$\overline{3}$m space group with two atoms per trigonal unit cell \cite{Cucka62} (Fig. \ref{trigonal}). 
Alternatively, the structure can be described as hexagonal with six atoms per unit cell (three primitive R3m unit cells) \cite{Hofmann2006}. In this structure, we can find two optical phonon modes. The first one, the $E_g$-mode, is a transverse optical mode doubly degenerate in the x-y-plane and consists of a displacement of the two atoms in opposite directions (Fig. \ref{trigonal} b)). The second one, the $A_{1g}$-mode, is a longitudinal optical mode, which consists of a displacement of the two atoms along the body diagonal in opposite directions (also referred to as breathing mode). Therefore, the incorporation of oxygen by thermal oxidation of Bi, or oxidation during any other growth methods will involve a modification of the structural properties affecting the lattice dynamics. These changes will hence be visible by crystal analysis, {\it e.g.}, by XRD or Raman scattering.

In this report we present results of our investigation of the structural and optical properties of Bi thin films grown onto amorphous SiO$_{2}$. Different thicknesses of Bi coatings are systematically studied by scanning electron microscopy (SEM), atomic force microscopy (AFM), and X-ray diffraction (XRD). These approaches reveal that Bi crystals grow preferentially in (111) orientation for film thicknesses starting from 20 nm and larger. Moreover, crystal properties and symmetry are analyzed by Raman Spectroscopy. This last contactless and nondestructive technique has arisen as an excellent tool for studying crystallinity, molecular interactions and phase and polymorphism, and allows to identify, with a high degree of certainty, the structure of Bi. It also gives evidence of possible incorporation of oxygen into the lattice, even if the layer is not fully covered by a single layer of oxygen. Surprisingly, our study suggests that Bi films preserve crystal quality, even for the thinner samples, by the observation of the optical phonons related to the trigonal unit cell of Bi. Furthermore, the absence of bismuth oxide features in the Raman spectra suggests that the surface remains free of oxidation.

\section{Experimental details}
The substrates whose nominal dimensions are 5 $\times$ 5 $ \times $ 1 mm, with  285  nm of SiO$_{2}$ on Si, were selected for growing Bi thin films. Before starting the process of growth, all the substrates were cleaned in ultrasonic bath sequences with acetone, ethanol, and finally isopropanol. The substrates were rinsed with MilliQ-water and then dried with pressurized nitrogen. 
Bismuth pellets with a purity of 99.99\%, having a melting point of  $\sim$  544 K, were placed in a tungsten crucible inside of a vacuum chamber and pumped to $5 \cdot 10^{-8}$mbar. The pellets were heated indirectly by passing a current through the crucible. The thermal evaporation deposited the bismuth on the silicon oxide substrate, which initially was at 298 K. During the growth process, the speed of evaporation was monitored by a quartz balance, and the deposition rate was kept at 0.1 \AA s$^{-1}$. Under these conditions, we grew samples with different thicknesses (5, 10, 20, and 40 nm). To analyze the grain size, shape, crystallinity, and oxidation, we have used SEM, AFM, XRD, and Raman spectroscopy.

We used SEM and AFM images to observe the shape of the grains and to measure their sizes in terms of the length of the edges surrounding them. We took the SEM images using an FEI Nova NanoSEM 200 instrument. Images were analyzed using the Image-J code \cite{ImageJ}. Topography and grain profiles were analyzed by tapping mode AFM using a Bruker Multimode V8 and Olympus brand silicon cantilevers with a nominal resonance frequency of 300 kHz and a spring constant of 26 N/m. WSM software \cite{ WsxM} was used for performing the data analysis. Crystallography characterization was done by XRD experiments in a Philips X'Pert-MRD instrument. The incident rays on the detector were attenuated for all the samples with a thickness equal to or bigger than 20 nm. The spectra obtained via this approach were compared with the established database at Ref. \cite{databasepeak}, which assigns crystallographic directions for every peak observed in the incident angle.

For Raman spectroscopy measurements at room temperature we used the 647 nm line of an Ar/Kr laser as excitation, a T64000 Jobin-Yvon spectrometer, coupled to an open electrode liquid-N$_{2}$ cooled Si-CCD detector, and an Olympus optical microscope. We used a $100\times$  optical objective (numerical aperture 0.95), both for focusing the laser beam, and  for collecting the scattered light. Optical characterization was performed using the triple spectrometer in subtractive mode, in order to achieve maximum resolution. Additionally, using the triple spectrometer, it is possible to go below 80 cm$^{-1}$ for the purposes of measuring the optical phonon modes of Bi. All measurements were calibrated with a Si  sample by the characteristic phonon peak at 519.5 cm$^{-1}$. After measuring the Si phonon peak, the spectrometer was not moved from its position in order to avoid displacements in the calibration; thus, our experimental values have a maximum error of 0.5 cm$^{-1}$. Special care was taken during Raman measurements to prevent thermal effects.  Specifically, the laser power was kept relatively low with long waiting times for data acquisition in order to avoid sample heating on the thinnest samples (5 nm). Finally, we apply Raman spectroscopy to analyze randomly selected points of the sample.

\section{Results and Discussion}
\subsection{Structural analysis by microscopy}

\begin{figure}[t!]
\centering
\includegraphics[width=0.8\linewidth]{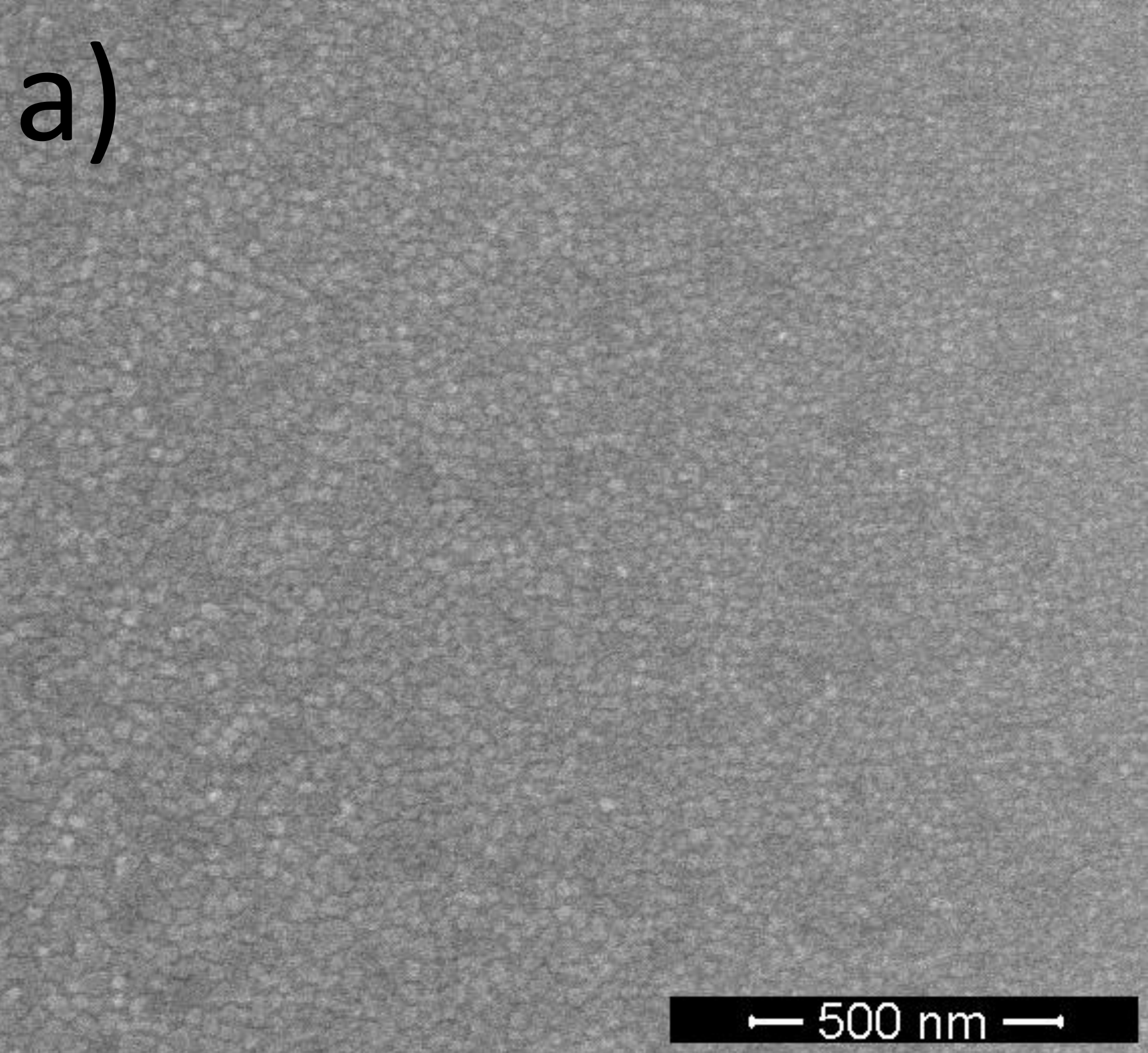}
\includegraphics[width=0.8\linewidth]{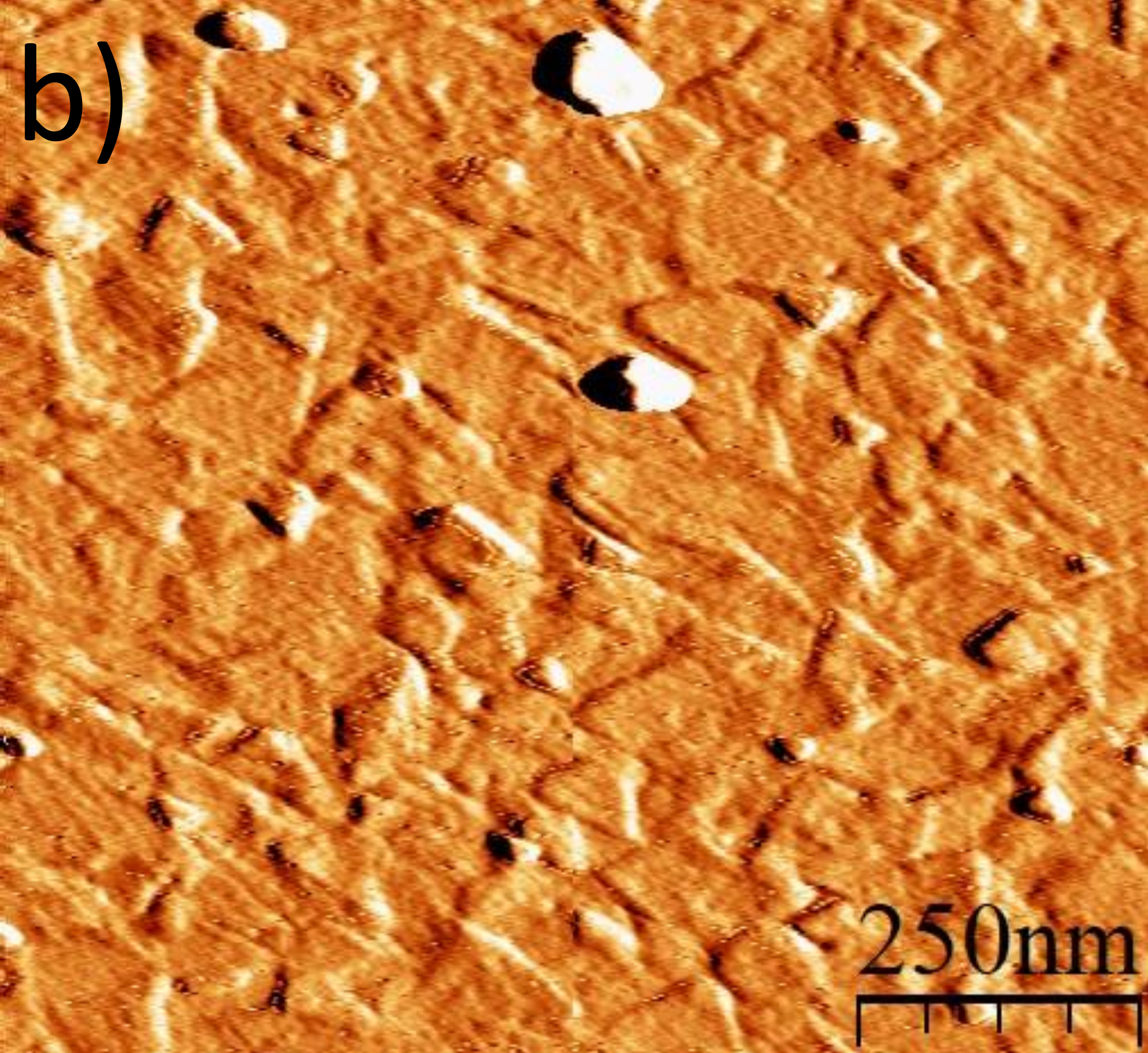}
\caption{Images of thermally grown Bi films on silicon oxide. a) SEM image for a 5 nm Bi film taken at 15 keV electron beam energy.  b) AFM image of the surface of a 40 nm Bi film. The scale of the images  is indicated at the lower right corner. The scales are similar, so that the images illustrate the great difference in grain sizes. }
\label{SEMAFM}
\end{figure}

The properties of Bi thin films are influenced by the thickness of the layer and by the substrate temperature during deposition \cite{biglass,morphologybi,KUMARI2007}. The deposition conditions strongly influence the structure and crystallography of the Bi layers \cite{Rodil2017,biglass,morphologybi,KUMARI2007}. We studied the morphology of the samples via SEM and AFM images. Homogeneous coating of the substrate and grain structures are illustrated in Fig. \ref{SEMAFM}. Panel a) displays a SEM image for a film of 5 nm Bi thickness; panel b) shows an AFM topographic image of a 40 nm thick Bi sample. Both panels have in common that Bi covers all the substrate homogeneously; however, they show that the grain sizes and shapes are completely different.

The SEM image of the panel a) shows tiny grains with different types of quadrilateral geometries, the most commonly observed are rhombuses, kites, and trapezoids. We have measured, independently of the shape, the perimeter length of the the edges of 30 different grains. The mean value and the standard deviation obtained for the edge lengths were $26 \pm 9$ nm.

Topographic images taken by AFM, Fig. \ref{SEMAFM} b), show a clear texture formed by geometrical triangular structures, that indicates the crystalline-like structure. Following the methodology employed for the panel \ref{SEMAFM} b), we calculated the mean value and standard deviation of the length of the edges for a selection of triangles, whose values are $100 \pm 40$ nm. This confirms the impression obtained from the images that the grain sizes for the 40 nm film are significantly larger than those for the 5 nm film. 

Our results obtained via SEM and AFM techniques are in good agreement with another studies \cite{Kumari}. The geometrical shapes observed for different thickness suggest the existence of crystallites on the amorphous substrates \cite{KUMARI2007}. However crystallographic studies and Raman should bring more information to confirm this point.

\subsection{Crystallographic analysis by XRD}

\begin{figure}[ht]
\centering
\includegraphics[width=1\linewidth]{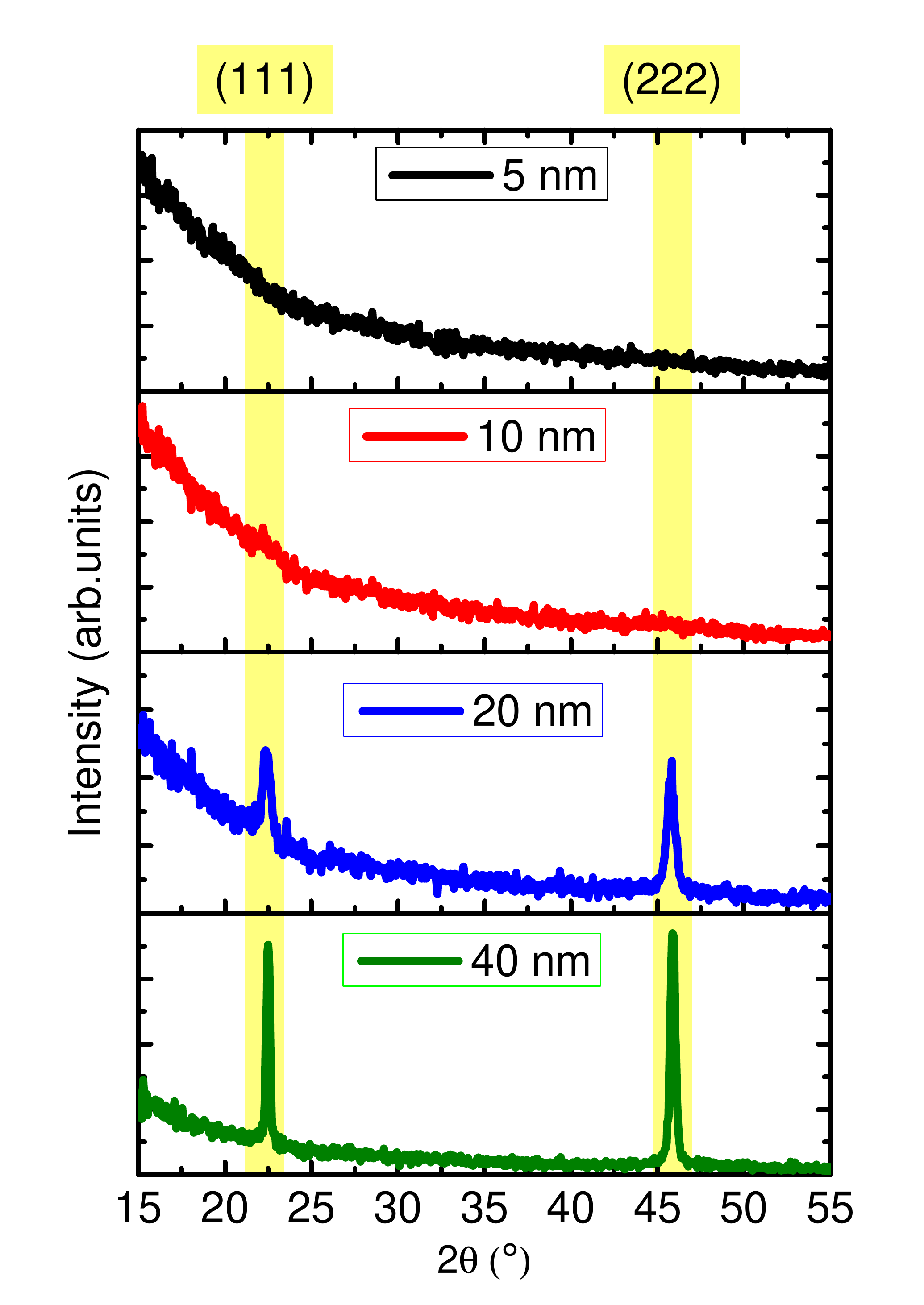}
\caption{X-ray diffraction spectra for Bi thin films on top of SiO$_2$. The four panels show XRD spectra for different thickness of Bi, as indicated in the legends.}
\label{DRXBi}
\end{figure}

The full coating of the Bi onto the SiO$_{2}$  and the appearance of the geometrical shapes (quadrilateral and triangular) suggest that the Bi films grow in a preferential crystallographic orientation. We have analyzed the crystallography data for all layer thicknesses. In Fig. \ref{DRXBi} we show in a stacked graph the X-ray intensity registered by the detector vs. the angle of incidence. The spectra are shown in four panels, with the green, blue, red, and black solid curves corresponding to 40, 20, 10, and 5 nm layer thickness, respectively.  As first observation, the spectra show more pronounced peaks (higher and narrower) when the Bi coating is thicker.

Figure \ref{DRXBi}  shows clear diffraction peaks corresponding to the positions of the (111) and (222) planes, supporting the interpretation of predominantly (111) textured crystallites. 
Going to 20 nm film thickness and above, the XRD spectra show peaks located at 22.5 and 45.8 degrees. According to  Ref. \cite{databasepeak}  the crystallographic (111) and (222) planes are located at 2$\theta = 22.5^{\circ}$ and 45.9$^{\circ}$, respectively. Yellow shades centered at 22.5 and 45.0 degrees help to visualize regions of the expected peaks for Bi (111) and (222). Moreover, the corresponding Miller index are shown in the upper parts of the yellow stripes.

The 5 nm thickness sample does not show a preferential crystallographic orientation, and it seems amorphous. In the panel for the 10 nm film a small protuberance around 22.5 degrees suggests the formation of crystallites in the (111) crystallographic directions. By subtraction of a smooth background we find that the evidence for the presence of a peak is fairly clear. 

The spectrum for the thinnest sample does not show characteristic peaks, which could induce us to think that the Bi layer is amorphous. However, Raman spectroscopy (see below) reveals the structure that is hidden by the XRD background signal of the silicon oxide substrate.
Table \ref{table:exp} summarizes the results of Fig. \ref{DRXBi}, and discussed in this section. 

\begin{center}
\begin{table}[ht]
\caption{\label{table:exp}Summary of the  position the peaks in 2$\theta$, for the spectra shown in Fig. \ref{DRXBi}.}
\begin{center}
    \begin{tabular}{|c|c|c|} 
    \hline
    \text{\bf Bi layer thickness} & \text{\bf $1^{st}$ Peak ($^{\circ}$)}&\text{\bf $2^{nd}$ Peak ($^{\circ}$)}   \\ \hline
    \textcolor{black}{5 nm} &\textcolor{black}{-}& \textcolor{black}{-}\\
    \textcolor{red}{10 nm} &\textcolor{red}{22.28}&\textcolor{red}{-} \\ 
    \textcolor{blue}{20 nm} &\textcolor{blue}{22.36} &\textcolor{blue}{45.82} \\    
    \textcolor{mygreen}{40 nm} &\textcolor{mygreen}{22.51}&\textcolor{mygreen}{45.87}\\ \hline
    \end{tabular}
\end{center}
\end{table}
\end{center}

\subsection{Raman spectroscopy}
The previous section leaves an open question on whether crystal symmetry is lost in the thinner samples when Bi is grown by thermal evaporation. To clarify this assertion, the structural and optical properties of these samples were evaluated by Raman spectroscopy.

\begin{figure}[htb]
\centering
\includegraphics[width=0.99\linewidth]{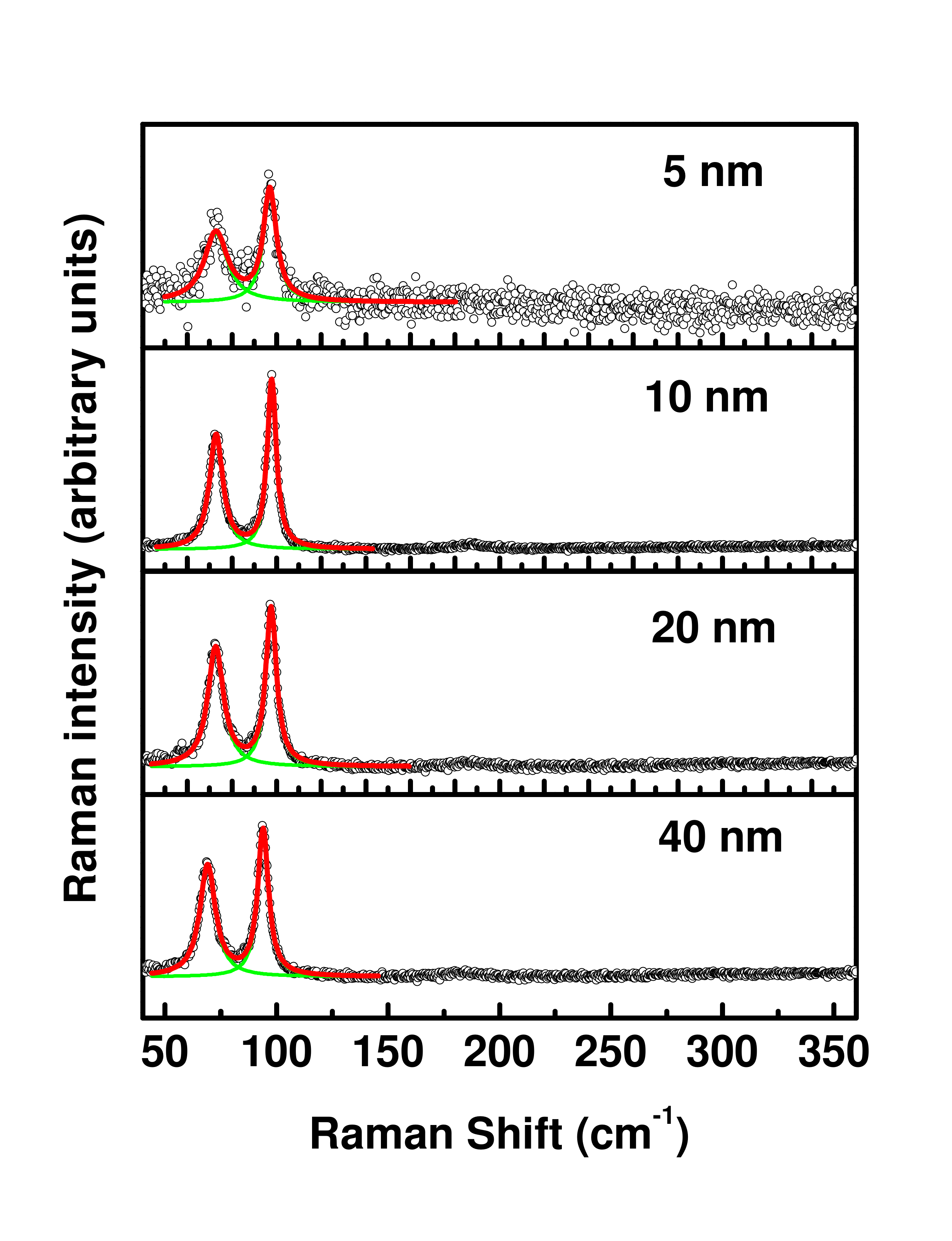}
\caption{Room-temperature Raman-scattering spectra of Bi thin films with a thickness of 5, 10, 20 and 40 nm (from the upper to lower panel). The green curves show Lorentzian line shapes fit to the spectra, for the two lines corresponding to the $E_g$ and $A_{1g}$ phonon modes of Bi. The red curve shows the cumulative fit, combining both lines.}
\label{Raman_new}
\end{figure}

The Raman spectra shown in Fig. \ref{Raman_new} were acquired for the four Bi thin films with different thicknesses (5, 10, 20, and 40 nm). 
Concerning the crystal structure of Bi, two clear peaks appear in all samples: the first-order Raman bands assigned to the $E_g$ and $A_{1g}$ phonon modes, in agreement with the literature \cite{Yarnell1964,Onari2002,Haro-Poniatowski1999}. Because of the centrosymmetric structure of the Bi lattice, these modes ({\it i.e.}, the $E_g$ and $A_{1g}$ gerade modes) do not show a dipole moment, but a change in the polarizability \cite{Rodriguez-Fernandez2016}. They are polarized in the hexagonal close-packed plane and, thus, in the rhombohedral structure. Compared with the sample of 40 nm thickness, where the phonon modes are centered at 69 and 94 cm$^{-1}$ (for $E_g$ and $A_{1g}$, respectively), we can appreciate a slight shift in the other samples, where phonons are located at 73 and 98 cm$^{-1}$. This blue-shift can be ascribed to phonon confinement due to the relaxation of  selection rules for reduced grain sizes \cite{Onari2002,Sahoo}. In particular, for the thinnest sample (5 nm), one can observe in the Raman spectra a broadening of the phonon peaks with a full width at half maximum (FWHM) of about 12 and 7 cm$^{-1}$ for the $E_{g}$ and $A_{1g}$ phonon mode, respectively. For Bi samples with thicknesses of 40, 20 and 10 nm, the FWHM of the $E_{g}$ and $A_{1g}$ band are about 7--8 cm$^{-1}$ and 4--5 cm$^{-1}$, respectively. This appears to be a significant result, since these are the clearest peaks found in the literature for pure thin films of Bi grown by thermal evaporation. Recent work reported Raman spectra for 1, 5, 10, and 30 nm Bi samples, which are comparable to this work \cite{Yang2019}. However, these samples were grown under UHV conditions by PLD, a technique that is widely employed for fabricating high-quality complex oxide films \cite{Zhang2012}. Therefore, while XRD does not provide information on Bi crystallinity for 10 and 5nm, we can appreciate that the crystallinity of the grains is preserved, based on the observation of the $E_g$ and $A_{1g}$ Raman modes for the thinnest layers.

\begin{center}
\begin{table}[htb]
\caption{\label{table1} Literature values of Raman frequencies for pure Bi and for some Bi$_{2}$O$_{3}$ phases. } 
\vspace{0.1cm}
\begin{center}

\begin{tabular}{cc}
\hline \hline
\multicolumn{2}{c}{\text{\bf Bi (rhombohedral structure)}}\\ 
Phonon modes&$\omega$ (cm$^{-1}$)
\\ \hline

$E_g$& $69$ (This work) / 70 \cite{Haro-Poniatowski1999}\\
$A_{1g}$& $94$ (This work) / 97 \cite{Haro-Poniatowski1999}  \\

\hline \hline
\\ \\
\hline \hline
\multicolumn{2}{c}{\text{\bf $\alpha,\beta,\gamma$-Bi$_{2}$O$_{3}$}}\\ 
Polymorphic&Raman peaks\\
phase&(cm$^{-1}$)\\ \hline
$\alpha$-Bi$_{2}$O$_{3}$& $65, 83, 93,102,118,138, 151,183$ \\
(monoclinic)& $210,279, 313, 410, 449...$\cite{Trentelman}\\
\hline
$\beta$-Bi$_{2}$O$_{3}$& $127,314,446$ \cite{Kumari}\\
(tetragonal)& \\
\hline
$\gamma$-Bi$_{2}$O$_{3}$& $58,89,145,166,208,250$ \\
(bcc)& $278,329,370,461...$\cite{Venu1972}\\
\hline \hline
\end{tabular}
\end{center}
\end{table}
\end{center}

It is worth highlighting that no peaks related to oxygen impurities are observed in the Raman spectra. It is known that Bi oxidizes more slowly according to the crystallographic facets due to the differing number of dangling bonds on each surface, which are the highest for $(10\overline{1})$ and lowest for $(111)$ \cite{Kowalczyk,JONA}. Accordingly, in the case of Bi, the $(111)$ facets are the most resistant to oxidation. The observation of the Raman phonon peaks for Bi $(111)$ suggests that $(111)$ facets remain the preferred orientation for 5 and 10 nm. In order to observe possible oxide impurities, we measured in the range of 40 to 360 cm$^{-1}$. For the purpose of this research, we have summarized the most characteristic Raman frequencies known for some Bi$_{2}$O$_{3}$ phases in  Table. \ref{table1}.

 The Raman scattering process involves the excitation of the material by a polarized laser beam. This means that the laser beam is focused to a spot size on the sample with a diameter of approximately 1--2 $\mu$m (using a 100$\times$ objective), where the absorption of radiation causes an increase in the temperature of the sample, which depends on thermal conductivity of the layer and substrate material. If the power intensity is highly increased, it can alter the lattice of the material, and result in breaking of the symmetry and loss of crystal quality; the material can even become amorphous  \cite{Shebanova}. This effect is more pronounced in thinner samples because thermal conductivity decreases as the size of the material scale down. Additionally, as in the case of Bi, the laser beam can induce oxidation and phase transitions \cite{Kumari}. The observation of the bulk Bi phonon lines and the absence of any bismuth oxide lines evidence that our precautions of using low laser power were sufficient to prevent alteration of the sample by laser irradiation.

\section{Conclusion}

In summary, we have grown Bi films of various thicknesses by thermal evaporation onto amorphous Silicon Oxide substrates. Morphological characterization by SEM and AFM shows that the amorphous substrate is fully coated by grains. Moreover, the geometrical structure of the grains was measured. For the thinnest sample, the shape of these grains was quadrilateral, and the mean value was $26 \pm 9$ nm, which suggests the existence of crystalline Bi. However, for the thicker layer, the shape was triangular with edges of $100 \pm 40$ nm, indicating a (111) orientation. We confirm this through XRD spectra, which demonstrate beyond doubt that the crystallographic direction of the grains is (111) for film thicknesses of 20 nm and higher. However, for 10 nm film thickness the data requires some fitting to elucidate the peak related to this orientation. In the case of 5 nm no signal was found in the XRD spectrum.

On the other hand, Raman spectroscopy measurements confirm the high crystallinity, even for the thinnest Bi films grown (thicknesses 5 and 10 nm). Moreover, we clarify that the Bi films are not strongly oxidized, even though they have been exposed to ambient conditions for more than one week. The Raman results contribute to the debate of the surface oxidation of Bi. Our results confirm that the crystallinity is preserved in the grains, even for the thinnest sample (5 nm), and give as predominant crystallographic direction the (111) orientation, which exposes the face that is most resistant to the oxidation. The results presented here for non-oxidized ultra-thin thermally evaporated Bi on SiO$_{2}$ offer new opportunities for obtaining  Bi (111) bilayers and support that this technique is as an excellent, simple and scalable to large areas for developing spintronics applications. Furthermore, we can take advantage of the work developed here, knowing that Bi bilayers are separated by a van der Waals gap forces \cite{Shu2016}. As a future work, we propose to explore electrochemical dissociation or polymer stamping methods for extracting Bi bilayers in future research.

\section{Acknowledgments}
CS and JMvR acknowledge financial support from  research programme of the Foundation for Fundamental Research on Matter (FOM), which is supported by the Netherlands Organization for Scientific Research (NWO). Also, CS gratefully acknowledges financial support offered by Generalitat Valenciana through PROMETEO2017/139 and GENT (CDEIGENT2018/028). We thank the Ministry of Finances and Competitiveness for its financial support through the grants RTI2018-093711-B-I00 and MAT2016-82015-REDT as well as FPI programme for young researchers.

\section{Author Information}
All authors have given approval to the final version of the manuscript.

\bibliographystyle{model1-num-names}
\bibliography{RamanSignalBitoSUBMIT.bib}

\end{document}